\documentstyle[12pt, epsf, bookman]{article}
\def\singlespace {\smallskipamount=3.75pt plus1pt minus1pt
                  \medskipamount=7.5pt plus2pt minus2pt
                  \bigskipamount=15pt plus4pt minus4pt
                  \normalbaselineskip=15pt plus0pt minus0pt
                  \normallineskip=1pt
                  \normallineskiplimit=0pt
                  \jot=3.75pt
                  {\def\smallskip {\vskip\smallskipamount}}
                  {\def\medskip   {\vskip\medskipamount}}
                  {\def\bigskip   {\vskip\bigskipamount}}
                  {\setbox\strutbox=\hbox{\vrule
                    height10.5pt depth4.5pt width 0pt}}
                  \parskip 7.5pt
                  \normalbaselines}
\def\middlespace {\smallskipamount=5.825pt plus1.5pt minus1.5pt
                  \medskipamount=11.25pt plus3pt minus3pt
                  \bigskipamount=22.5pt plus6pt minus6pt
                  \normalbaselineskip=22.5pt plus0pt minus0pt
                  \normallineskip=1pt
                  \normallineskiplimit=0pt
                  \jot=5.825pt
                  {\def\smallskip {\vskip\smallskipamount}}
                  {\def\medskip   {\vskip\medskipamount}}
                  {\def\bigskip   {\vskip\bigskipamount}}
                  {\setbox\strutbox=\hbox{\vrule
                    height15.75pt depth6.75pt width 0pt}}
                  \parskip 7.25pt
                  \normalbaselines}
\def\dblspc {\smallskipamount=7.5pt plus2pt minus2pt
                  \medskipamount=15pt plus4pt minus4pt
                  \bigskipamount=30pt plus8pt minus8pt
                  \normalbaselineskip=30pt plus0pt minus0pt
                  \normallineskip=2pt
                  \normallineskiplimit=0pt
                  \jot=7.5pt
                  {\def\smallskip {\vskip\smallskipamount}}
                  {\def\medskip   {\vskip\medskipamount}}
                  {\def\bigskip   {\vskip\bigskipamount}}
                  {\setbox\strutbox=\hbox{\vrule
                    height21.0pt depth9.0pt width 0pt}}
                  \parskip 15.0pt
                  \normalbaselines}
\def\nb{\nabla }
\def\gm{\gamma }
\def\al{\alpha }

\def\be{\begin{equation}}

\def\j-{\J_-}

\def\ee{\end{equation}}

\def\bearr{\begin{eqnarray}}
\def\bearrs{\begin{eqnarray*}}
\def\eearr{\end{eqnarray}}
\def\eearrs{\end{eqnarray*}}
\def\barr{\begin{array}}
\def\earr{\end{array}}
\def\p{\partial}

\def\th{\theta}

\def\non\non{\nonumber}
\def\nn8{\nonumber\\[15pt]}
\def\l{\left}
\def\r{\right}
\def\un{\underline}

\def\f{\frac}

\def\dis{\displaystyle}

\begin{document}
\input epsf
\def \figdraw #1 {        \newbox\boxtmp
        \setbox\boxtmp=\hbox{\epsfbox{#1}}
        \usebox{\boxtmp} \\}
\def \putname #1 {\vfil \rightline{#1}}
\middlespace
\begin{center}
{\Large{\bf Inertial Forces as Viewed from the\\
ADM Slicing and Their Behaviour for Particles\\
in Non-Circular Geodesics}}\\[30pt]
A.R.Prasanna\\

Physical Research Laboratory\\
 Ahmedabad 380 009,
India\\
(email: prasanna@prl.ernet.in)\\[12pt]
\end{center}
\middlespace
\begin{center}
\un{Abstract}\\
\end{center}

Considering the definition of inertial forces acting on a test
particle, following non-circular geodesics, in static and
stationary space times we show that the centrifugal force reversal
occurs only in the case of particles following prograde orbits
around black holes. We first rewrite the covariant expressions for
the acceleration components in terms of the lapse function, shift
vector and the 3-metric $\gm_{ij}$, using the ADM 3+1 splitting
and use these, for different cases as given by pure radial motion,
pure azimuthal motion and the general non-circular motion. It is
found that the reversal occurs only when the azimuthal angular
velocity of the particle supersedes the radial velocity, which
indeed depends
upon the physical parameters $E$, $\ell$ and the Kerr parameter $a$.\\[50pt]



\newpage

\section{Introduction}
Though the idea of introducing the concept of inertial forces
within the framework of general relativity is more than a decade
old (Abramowicz et al., 1988, 1990, 1993), one of the significant
results of doing so, viz. the reversal of centrifugal force
(Abramowicz and Prasanna, 1990) seems to have found some
applicability in the context of astrophysical scenario recently.
Heyl (2000) while probing the properties of neutron stars with
type 1 X-ray bursts, concludes that, if the change in spin
frequency is due to a change in the thickness of the atmosphere,
the radius of the star must exceed $3m$ for any equation of state,
the constraint arising from the reversal of centrifugal force. He
also considers the effect of the Coriolis type of force arising
from the dragging of inertial frames and its relevance in the
context of calculating the angular momentum of the fluid element
in the neutron star atmosphere. More recently, Hasse and Perlick
(2001) while discussing gravitational lensing in spherically
symmetric static space times with centrifugal force reversal, show
that in any spherically symmetric static space time, the upper
limiting radius for a non-transparent Schwarzschild source to act
as a gravitational lens that produces infinitely many images, is
intimately related to the radius at which the centrifugal reversal
takes place. While these studies are encouraging to show the
relevance of the introduction of `inertial forces' in general
relativity, one has to bear in mind that almost all the results
obtained earlier in the context of `centrifugal reversal' were for
the study of test particles in circular geodesics. In fact, Gupta,
Iyer and Prasanna (1996) had obtained the expression for the
centrifugal force on a fluid element, in the optical reference
geometry approach while considering the behaviour of ellipticity
of a slowly rotating configuration which is ultra compact, but the
treatment was not fully covariant.\\

We shall now set up a formalism using the covariant prescription
for inertial forces (Abramowicz et al. (1995) and relating it to
the study of test fluids on a given background geometry in the
ADM splitting with shift vector and lapse function and rewrite
the components of acceleration in terms of these quantities and
the velocity 3-vector $V^i$. With this formalism, one can
perhaps look for the behaviour of the inertial forces in general
dynamical space times using numerical techniques.\\

In the present paper, we shall use the formalism for considering
fluid flow in stationary space times with only radial and only
azimuthal components of velocity being non-zero and then the
case of static space time wherein both the radial and azimuthal
components of particle three velocity are non-zero (non-circular
godesics).
\section{Formalism}
On a general manifold M described by the line element
\be
ds^2 = g_{\al \beta} dx^\al dx^\beta \ee One can use the 3+1 ADM
splitting to define the lapse function $\al$, shift vector
$\beta^i$ and the 3-metric $\gm_{ij}$ (York 1983) to rewrite (1)
as
\be
ds^2 = - \l( \al^2 - \beta_i \beta^i \r) dt^2 + 2\beta_i dx^i dt
+ \gm_{ij} dx^i dx^i
\ee
It is well known that this splitting forms the basis for
numerical hydrodynamics, wherein with the introduction of a unit
time-like vector field $n^\mu$ normal to the space-like
hypersurface $\sum$ ($t$ = const), one can define the fluid
3-velocity $V^i$ (York 1983)
\be
V^i = \f{U^i}{\al U^t} + \f{\beta^i}{\al}
\ee
with $U^i$ being the spatial component of the fluid four
velocity $U^\mu = \l( U^t, U^i \r)$.\\

In order to introduce the ``inertial forces" (Newtonian) one can
follow the covariant procedure as given by Abramowicz (1995), and
relate the time-like vector field $n^\mu$ to the gradient of the
gravitational potential $\phi$ through
\be
n^\nu \nb_\nu n_\mu = - \nb_\mu \Phi \ee with
\be
\phi = \f{1}{2} \ell n \l[ - < \eta , \eta > - 2\omega < \xi ,
\eta > - \omega^2 < \xi \xi > \r] \ee for any stationary
axisymmetric space-time with $\eta$ and $\xi$ denoting the
time-like and space-like Killing Vectors.\\

If $\tau^\mu$ represents a space-like vector field, orthogonal to
$n^\mu$, then the four velocity in the space-time may be written
as
\be
\barr{lll} U^\mu&=& \gm \l( n^\mu + V \tau^\mu \r)\\[16pt] \gm&=&
n^\mu U_\mu = \f{1}{\sqrt{1-V^2}} \earr \ee and the three metric
\be
\barr{lll} \gm_{ij}&=& g_{ij} + n_i n_j\\[16pt] \gm_{ij} V^iV^j&=&
V^2 \earr \ee If we now introduce the conformal rescaling of the
3+1 splitting as done by Abramowicz, Carter and Lasota (1988)
using what is called the Optical Reference Geometry
\be
\barr{lll}
\tilde{\gm}_{ik} &=& e^{2\phi} \gm_{ik}\\[16pt]
\tilde{\tau}^i&=& e^{-\phi} \tau^i
\earr
\ee
then the four acceleration $a_\mu = U^\nu \nb_\nu U_\mu$ can be
written in the explicit form (Abramowicz 1993)
\be
\barr{lll} a_\mu :&=& - \nb_\mu \phi + \gm^2V \l( n^\nu \nb_\nu
\tau_\mu + \tau^\nu \nb_\nu n_\mu \r) + \dot{\gm} n_\mu\\[16pt] &&+
\tilde{V}^2\tilde{\tau}^\nu \tilde{\nb}_\nu \tilde{\tau}_\mu + \l(
\dot{E} V e^\phi + \gm \dot{V} \r) \tau_\mu \earr \ee wherein the
terms can get the specific identifications as gravitatonal,
Coriolis type (Lense-Thirring) $\th (V)$, centrifugal $\th \l( V^2
\r)$ and Eulerian accelerations. $\tilde{\nb}_\nu$ is the
covariant derivative in the absolute 3-space with the projected
conformal metric $\tilde{\gm}_{ij}$.\\

One can now relate these two splittings of the axisymmetric,
stationary space-time through the definition
\be
\barr{lll} n^\mu&=& \l( \f{1}{\al}, - \f{\beta^i}{\al} \r)\\[16pt]
n_\mu&=& \l( - \al , 0 \r)\\[16pt] \tau^\mu&=& \l( 0, \f{V^i}{V}
\r)\\[16pt] \tau_\mu &=& \l( \f{\beta^iV_i}{V} , \f{V_i}{V} \r)
\earr \ee Using (9) and (10) one can express the components of
acceleration vector $a_\mu$ in terms of $\al$, $\beta^i$,
$\gm_{ij}$ and the 3-velocity components $V^i$.\\

The centrifugal acceleration acting on a fluid element
$\tilde{V}^2 \tilde{\tau}^\nu \tilde{\nb}_\nu \tilde{\tau}_\mu$ is
given by
\be
(Cf)_i= \gm^2 \l[ V V^j \p_j \l( \f{V_i}{V} \r) + \l( V_i V^j \p_j
- V^2 \p_i \r) \l( \Phi \r) - \f{1}{2} V^j V^k \p_i \gm_{jk} \r]
\ee while the Coriolis type (Lense-Thirring) is given by
\be
(Co)_i = - \f{\gm^2}{\al} \l[ V\beta^j\p_j \l( \f{V_i}{V} \r) +
V^j \p_i g_{oj} - \beta^k V^j \p_i \gm_{kj} \r] \ee Thus for any
given background geometry one can evaluate the specific
acceleration components, if one has the 3-velocity field of the
fluid on that geometry evaluated through the equations of motion.
\section{Specific Examples}
\subsection{Purely Radial Flow $\l(  V^r \neq 0, V^\phi = 0,
V^\th = 0 \r)$} A purely radial flow which is encountered in the
case of spherical accretion, particularly in static space times
will have
\be
(Cf)_\gm = \gm^2 \l[V V^r \p_r \l( \f{V_r}{V} \r) -\f{1}{2} \l(
V^r \r)^2 \p_r \gm_{rr} \r] \ee As $V^2 = \gm_{rr} \l( V^r\r)^2$
and $V_r = \gm_{rr}V^r$, the first term also gives $\f{1}{2} \l(
V^r\r)^2 \p_r\gm_{rr}$ and thus $(Cf)_r$ is identically zero. This
is to be expected and goes to show the consistency of the
definition.
\subsection{Purely Azimuthal Flow $\l(  V^r = 0, V^\th = 0,
V^\phi \neq 0 \r)$}
\be
\barr{lll} (Cf)_r&=& \gm^2 \l[ V V^r \p_r \l( \f{V_r}{V} \r) +
\l( V^r V_r \p_r - V^2 \p_r \r) \l( \Phi \r) - \f{1}{2} V^j V^k
\p_r \gm_{jk} \r]\\[16pt] &=& \gm^2 \l[ - \gm_{\phi\phi} \l(
V^\phi\r)^2 \l( \p_r \Phi \r) - \f{\l( V^\phi \r)^2}{2} \p_r
\gm_{\phi\phi} \r]\\[16pt] (Cf)_\gm&=& - \f{\gm^2\l(
V^\phi\r)^2}{2} \l[ 2 \gm_{rr} \l( \p_r \Phi \r)+ \p_r
\gm_{\phi\phi} \r]  \earr \ee

~~\\
a.~~\un{Schwarzschild Space Time}:\\
\be
\barr{lll} \Phi&=& - \f{1}{2} ln \l( 1 - \f{2m}{r} \r)\\[16pt]
\gm_{\phi\phi}&=& r^2\\[16pt] (Cf)_r&=& - \gm^2 \l( V^\phi\r)^2 r
\l( 1 - \f{2m}{r} \r)^{-1} \l( 1 - \f{3m}{r} \r) \earr \ee
\newpage
b.~~\un{Kerr Space Time}:\\
\be
\barr{lll} \Phi&=& - \f{1}{2} ln \l\{ \f{\dis \l( r^3 + a^2 r -
2mr^2 \r)}{\dis \l( r^3 + a^2r + 2ma^2 \r)} \r\}\\[16pt]
\gm_{\phi\phi}&=& r^2 + a^2 + \f{2ma^2}{r}\\[16pt] \Delta&=& r^2 -
2mr + a^2\\[16pt] (Cf)_r&=& \f{\gm^2 \l( V^\phi\r)^2}{r^2\Delta}
\l( r^5 - 3mr^4 + a^2 \l( r^3  - 3mr^2 + 6m^2r - 2m^2 \r) \r)
\earr \ee
The results in this case is again as expected and is the same as
in the test particle case (Abramowicz and Prasanna (1990), Iyer
and Prasanna (1993)).\\

Case (3): $V^r \neq 0$, $V^\phi \neq 0$, $V^\th = 0$.\\
\begin{itemize}
\item[(a)]
As mentioned earlier, the discussion of inertial forces was
earlier confined to particles in circular motion only. Hence, it
would be useful first to consider the case of a test particle with
both radial and azimuthal velocity non-zero. We shall restrict the
discussion for static geometry as given by \be ds^2 = - \l( 1 -
\f{2m}{r} \r) dt^2 + \l( 1 - \f{2m}{r} \r)^{-1} dr^2 +
r^2d\Omega^2 \ee For a test particle following a geodesic,
confined to the equatorial plane $\th = \f{\pi}{2}$, the equations
of motion reduce to \be \l( 1 - \f{2m}{r} \r) \f{dt}{ds} = E \ee
\be r^2 \f{d\phi}{ds} = \ell \ee and \be \f{dr}{ds} = \sqrt{E^{2}
-  \l( 1 - \f{2m}{r} \r) \l( 1 + \f{\ell^2}{r^2} \r)} \ee From the
definition of the 3-velocity $V^i = \f{u^i}{\al u^o}$, we then get
\be \l( V^r \r)^2 =  \l( 1 - \f{2m}{r}  \r) \l[ 1 - \f{1}{E^2} \l(
1 - \f{2m}{r}\r) \l( 1 + \f{\ell^2}{r^2} \r)\r] \ee and \be \l(
V^\phi \r)^2 = \f{\ell^2}{E^2 r^4} \l( 1 - \f{2m}{r} \r) \ee
Using
these in the expression for the centrifugal force, one finds
\be
Cfr = \f{- m \ell^2}{r^3 \l[ 2m - \l( 1 - E^2 \r) r \r]}
\ee
which clearly shows no reversal for any value of $E$ and $\ell$,
outside the horizon.\\
\item[(b)] We shall consider a second example of static spacetime,
but with an additional interacting field namely that of a charged
particle in the Reissner-Nordstrom geometry.
\[
\barr{lll}
ds^2&=& - \l( 1 - \f{2m}{r} + \f{Q^2}{r^2} \r) dt^2 - \l( 1 -
\f{2m}{r} + \f{Q^2}{r^2} \r)^{-1} dr^2\\[10pt]
&&- r^2 d\th^2 - r \sin^2\th d\phi^2
\earr
\]
For a charged test particle with charge $e$ one has the energy $E$
and angular momentum $\ell$ as given by
\[
\barr{lll}
U^t&=& \l( 1 - \f{2m}{r} + \f{Q^2}{r^2} \r)^{-1} \l( E - \f{eQ}{r} \r)\\[12pt]
U^\phi&=& \f{\ell}{r^2}
\earr
\]
which in turn give for the components of three velocity
\[
\barr{lll}
\l( V^r \r)^2&=& \l( 1 - \f{2m}{r} + \f{Q^2}{r^2} \r) \l\{ 1 - \l( 1
- \f{2m}{r} + \f{Q^2}{r^2} \r) \l( 1 + \f{\ell^2}{r^2} \r) / \l( E -
\f{eQ}{r} \r)^2 \r\}\\[12pt]
\l( V^\phi \r)^2&=& \f{\ell^2}{r^4} \l( 1 - \f{2m}{r} + \f{Q^2}{r^2}
\r) / \l( E - \f{eQ}{r} \r)^2
\earr
\]
Evaluating the centrifugal force acting on such a particle one finds
\[
\l( Cfr \r) = \f{\ell^2 \l[ \l( 1 - e^2 \r) Q^2 - mr + eEQr\r]}{r^3
\l\{ \l( 1 - e^2 \r) Q^2 - 2eEQr +r \l( 2m - \l( 1 - e^2 \r) r \r)
\r\}}
\]
It is easy to see that $Cfr$ can be zero only for $r = Q^2 \l( 1 -
e^2 \r) / (m-eEQ)$, which is $<m$ and thus inside the event horizon.
\item[(c)] Stationary Space-time:\\

We consider the case of a test particle in a non-circular orbit
($V^r$, $V^\phi$, both non-zero) on the equatorial plane of the Kerr
geometry which represents the space-time outside a black hole as
given by
\[
ds^2 = - \l( 1 - \f{2m}{r} \r) dt^2 + \f{r^2}{\Delta} dr^2 +
\f{4am}{r} d\phi dt + \l( r^2 + a^2 + \f{2ma^2}{r} \r) d\phi^2
\]
The constants of motion $E$ and $\ell$ of the particle are now
related as given by
\[
\barr{lll}
U^\phi&=& \f{d\phi}{ds} = \f{1}{\Delta} \l[ \l( 1 - \f{2m}{r} \r)
\ell + \f{2am}{r} E \r]\\[12pt]
U^t&=& \f{dt}{ds} = \f{1}{\Delta} \l[ \l( r^2 + a^2 + \f{2ma^2}{r}
\r) E - \f{2ma}{r} \ell \r]
\earr
\]
and the radial four velocity $U^r$ is given by
\[
\l( U^r \r)^2 = \l( \f{dr}{ds} \r)^2 = E^2 + \f{2m}{r^3} \l( a E -
\ell \r)^2 + \f{1}{r^2} \l( a^2 E^2 - \ell^2 \r) - \f{\Delta}{r^2}
\]
The lapse function $\al$, the shift vector $\beta^i$ and the 3-metric
$\gm_{ij}$ are given by
\[
\barr{lll}
\al^2&=& \f{\Delta}{\l( r^2 + a^2 + \f{2ma^2}{r}\r)}\\[12pt]
\beta^i&=& \l( 0, \; 0, \; \f{-2am}{\l( r^3 + a^2 r + 2ma^2 \r)} \r)
\earr
\]
and
\[
\gm_{ij} = \l[ \barr{ccc} \f{r^2}{\Delta}&0&0\\
0&0&0\\
0&0&\l( r^2 + a^2 + \f{2ma^2}{r} \r) \earr
\r]
\]
From these expressions one can easily obtain the radial and azimuthal
velocities of the particle in the 3-space as given by
\[
\barr{lll}
V^r&=& \f{U^r}{\al U^t}\\[10pt]
V^\phi&=& \f{\l(  U^\phi / U^t + \beta^\phi \r)}{\al}
\earr
\]
Using these in the expression for the `Centrifugal Acceleration'
$(Cf)r$ as given by (11), one finds
\[
\l( Cf \r)_r = \f{\barr{l} - \l[ \ell^2 mr \l\{ r^7 - 6aE\ell r^5
+ \l( r + 2m \r) \l( a^6 - 2a^5 E \ell \r) \right. \right.\\
  - 4 a^3 E \ell r^2 \l( 3m + 2r \r) + a^4 \l( 4 m\ell^2 - 8 m^2 r
+ 3r^3 \r)\\
\left. \left. + a^2 \l( 12 \ell^2 mr^2 + r^4 \l( 3r - 2m \r) \r)
\r\} \r] \earr}{\barr{l}\l[ \l( r^3 + a^2 \l( 2m + r \r)
\r)^2 \l( - 4a E\ell mr^3 - 4a^3 E\ell m \l( r + 2m \r) + \right. \right.\\
r^5 \l( 2m + \l( E^2 - 1 \r) r \r) + a^4 \l( r + 2m \r) \l( E^2
\l( r + 2m \r) - r \r)\\
 \left. \left. + 2a^2 \l( 2\ell^2 m^2 + r^2 \l( 2m^2 + 2E^2 mr + \l( E^2
- 1 \r) r^2 \r) \r) \r) \r] \earr }
\]
If one looks for solutions of the equation for $\l( Cf\r)_r = 0$, in
principle, one finds eight roots of which there may be just one or
two outside the horizon $r_+ =m + \sqrt{m^2-a^2}$, for different
combinations of $E$ and $\ell$. However, if one also demands the
physicality of the root by looking at the condition $\l( U^r \r)^2 >
0$, then it appears that there exists one root between $2m < r < 3m$,
and the exact location depends upon the combination of $E$ and
$\ell$ of the particle. One can also see that the location of the
realistic root for $\l( Cf \r)_r = 0$ appears after the point wherein
$\f{U^\phi}{U^t}$ exceeds $\f{U^r}{U^t}$ for the particle moving
inwards from infinity.
\end{itemize}
\section{Discussion}
The reversal of centrifugal force acting on a test particle in
circular orbit at the last circular unstable photon orbit, in
static space time has
drawn some attention in recent times in
varying context ranging from X-ray sources to infinitely multiple
image forming. Actually one can clearly see that these features
are closely associated with the photon behaviour in static space
times. Though test particle trajectories do give some
understanding of the geometry on which they are moving, for
astrophysical applications it is fluid flows that are important.\\

Having obtained the expressions for the inertial accelerations
in terms of the fluid velocity components and the metric
components one can now write down the individual forces on any
background space time, provided one has the solution for the
velocity components on the given background for the flow.
For a dusty fluid ($p = 0$) as may be expected for
a purely radial flow $\l( V^\th = 0, V^\phi = 0 \r)$ the
centrifugal acceleration is zero, whereas for a purely azimuthal
flow $\l( V^r = 0, V^\th = 0 , V^\phi \neq 0 \r)$, the
centrifugal acceleration is just as for a test particle, both in
Schwarzschild and the Kerr background. The fluid in this case is
a collection of test particles in circular orbits and thus the
result is as known earlier.\\

However, it is important to understand the behaviour of forces for
a more general motion of particles, when the trajectory is a
non-circular geodesic. For such a case, it is easy to find the
expressions for three velocity components from the components of
the four velocity, which are directly integrable from the
equations of motion using the symmetries of the given space time.
This would bring in the physical characteristics of the particle,
the energy $E$ and angular momentum $\ell$ and restricting the
discussion to particles on the equatorial plane $\l( \th =
\f{\pi}{2}, \; U^\th = 0 \r)$, one can easily obtain the
components $V^r$ and $V^\phi$ for the particle.\\

From the expression for $\l( Cf \r)_r$ in Kerr space time, it is
clear that its behaviour depends upon the particle parameters $E$
and $\ell$ and the rotation parameter `$a$'. Direct evaluation for
fixed $E$ and $\ell$ shows that when $E \leq 1$, the reversal
radius occurs outside the event horizon only for very high value
of $a$ (Table 1). As may be seen, the table also gives for the
same set of parameters the radius at which $\l( F_3 =
\f{U^\phi}{U^t} \r)$ crosses over $\l( F_2 = \f{U^r}{U^t} \r)$,
and it is clear that the centrifugal reversal occurs only after
the angular velocity supersedes the radial velocity. Figures 1 to
3 show the behaviour of the curves, centrifugal $\l( f1 \r)$,
radial velocity $\l( f2 \r)$ and the
azimuthal velocity $\l( f3 \r)$ for different values of $a$, for same
$E$ and $\ell$.\\

\middlespace Indeed one can see that the centrifugal reversal
which was inherent for circular geodesics in both static and
stationary spacetimes, does not follow automatically for general
non-circular geodesics. Whereas it does not occur at all in static
spacetime, in stationary spacetime, the occurrence depends upon
the energy, angular momentum of the particle and the rotation
parameter $a$, which needs to be sufficiently high for the
reversal to occur outside the event horizon. The difference in the
behaviour for static and stationary spacetimes is essentially due
to the fact that the rotation induces `frame dragging', which adds
to the azimuthal velocity and makes it larger than the radial
velocity after which the behaviour resembles that of a circular
orbit. This interpretation gets further support from the fact that
for particles in retrograde motion $\l( a > 0, \; \ell < 0 \r)$,
there are no positive real roots for the equation $\l(
Cf \r)_r = 0$.\\

Thus we find that for particles on non-circular geodesics, in
static spacetime, there is no reversal of centrifugal force, and
in Kerr spacetime, the reversal occurs only for particles in
prograde motion, which get the additional input to their angular
velocity from the effects of frame dragging. For particles in
retrograde motion, the frame dragging contribution would not
suffice to overcome the effects of radial velocity and thus like
in static case shows no reversal.\\
\newpage
\singlespace
\begin{center}
\un{Table 1}\\[12pt]

Location of the last root of $CFr = 0$ and the second root of
$(F2-F3)=0$,\\ alongwith the location
of  event horizon $(EH)$ for different values of $a$\\[12pt]

\begin{tabular}{||c|c|c|c||}
\hline\hline
$a$&$EH$&$(Cfr) =0$&$(F2-F3)=0$\\
\hline
0.1&1.99499&----&2.0615\\
\hline
0.2&1.9798&----&2.10945\\
\hline
0.3&1.95394&----&2.14586\\
\hline
0.4&1.91652&1.28011&2.17201\\
\hline
0.5&1.86603&1.48094&2.18871\\
\hline
0.6&1.8&1.54586&2.19647\\
\hline
0.7&1.71414&1.78834&2.19554\\
\hline
0.8&1.6&1.91397&2.18595\\
\hline
0.9&1.43589&2.02589&2.16752\\
\hline
1.0&1.0&2.12612&2.13987\\
\hline\hline
\end{tabular}
\end{center}

\vspace*{1in}

\begin{center}
\def \epsfsize#1#2{0.6#1}
\figdraw{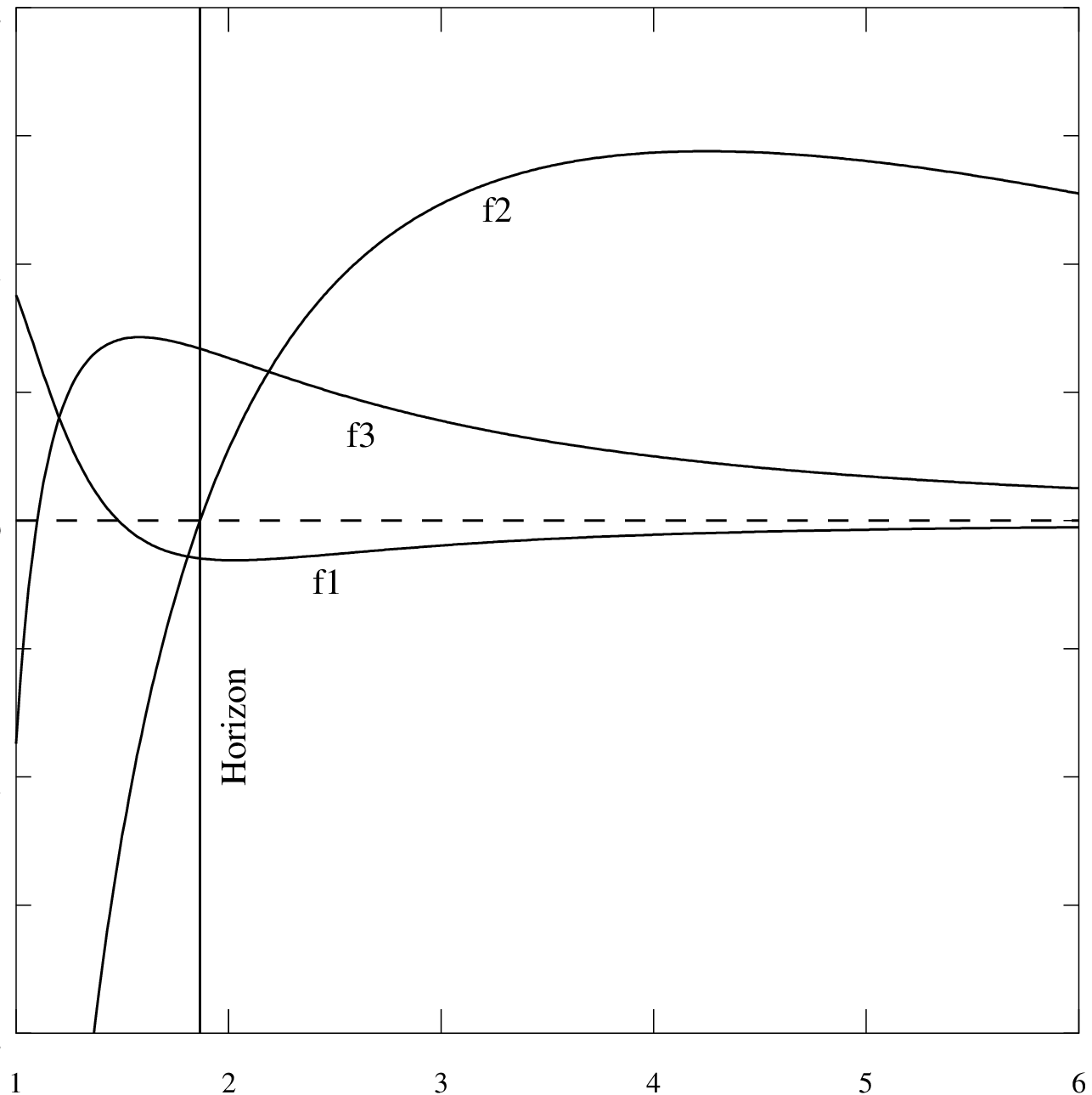}
\end{center}
\singlespace \noindent {\footnotesize Fig. 1:
Plots of $Cfr \l( f1 \r)$, $\f{U^r}{U^t} \l( f2 \r)$ and
$\f{U^\phi}{U^t} \l( f3 \r)$ for the parameter values $E = 0.9$,
$\ell =1$ and $a=0.5$. The vertical line shows the location of the horizon.
}

\newpage
\begin{center}
\def \epsfsize#1#2{0.6#1}
\figdraw{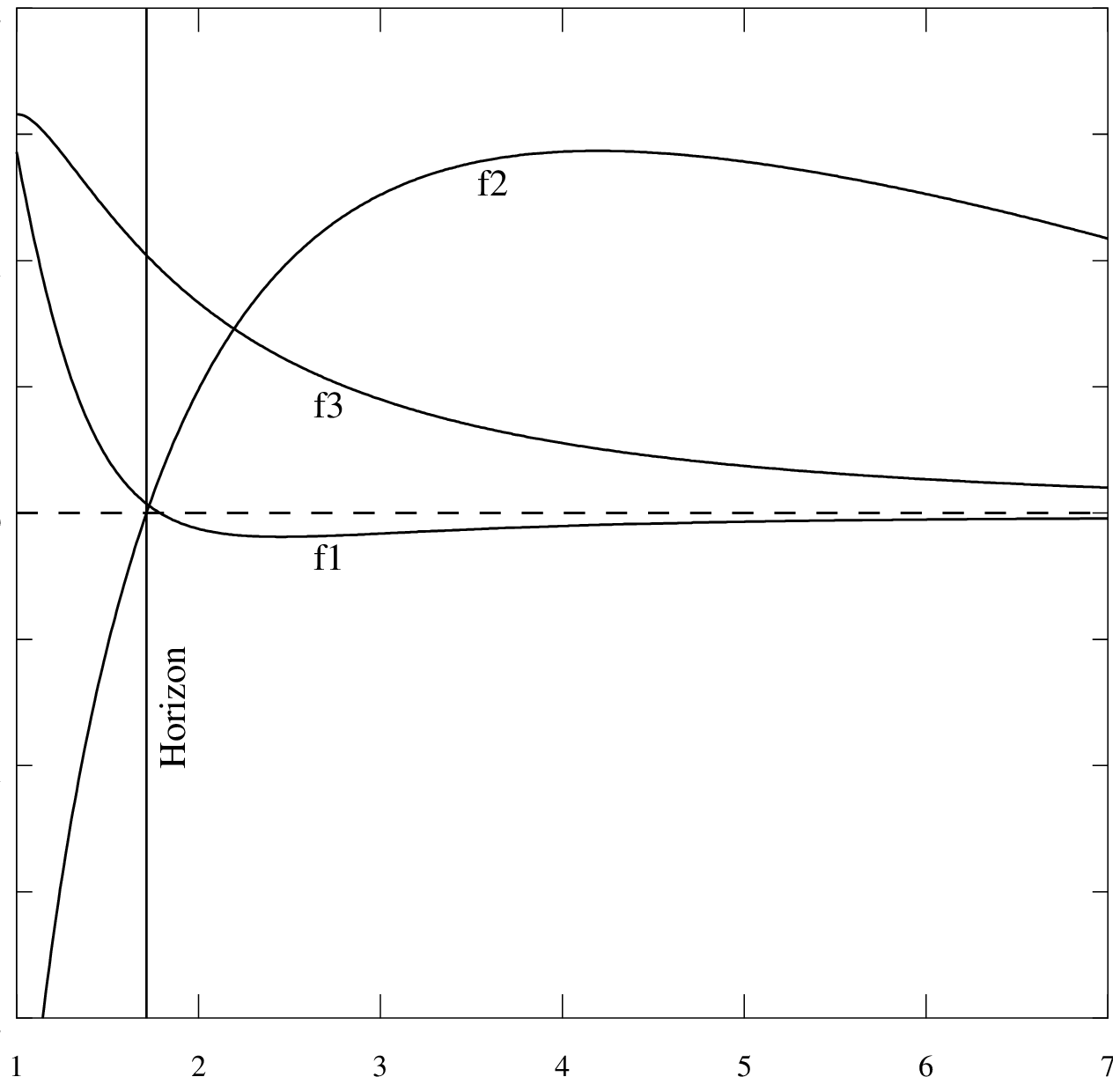}
\end{center}
\singlespace \noindent {\footnotesize Fig. 2:
Same as Fig. 1, but for $a = 0.7$.
}

\begin{center}
\def \epsfsize#1#2{0.6#1}
\figdraw{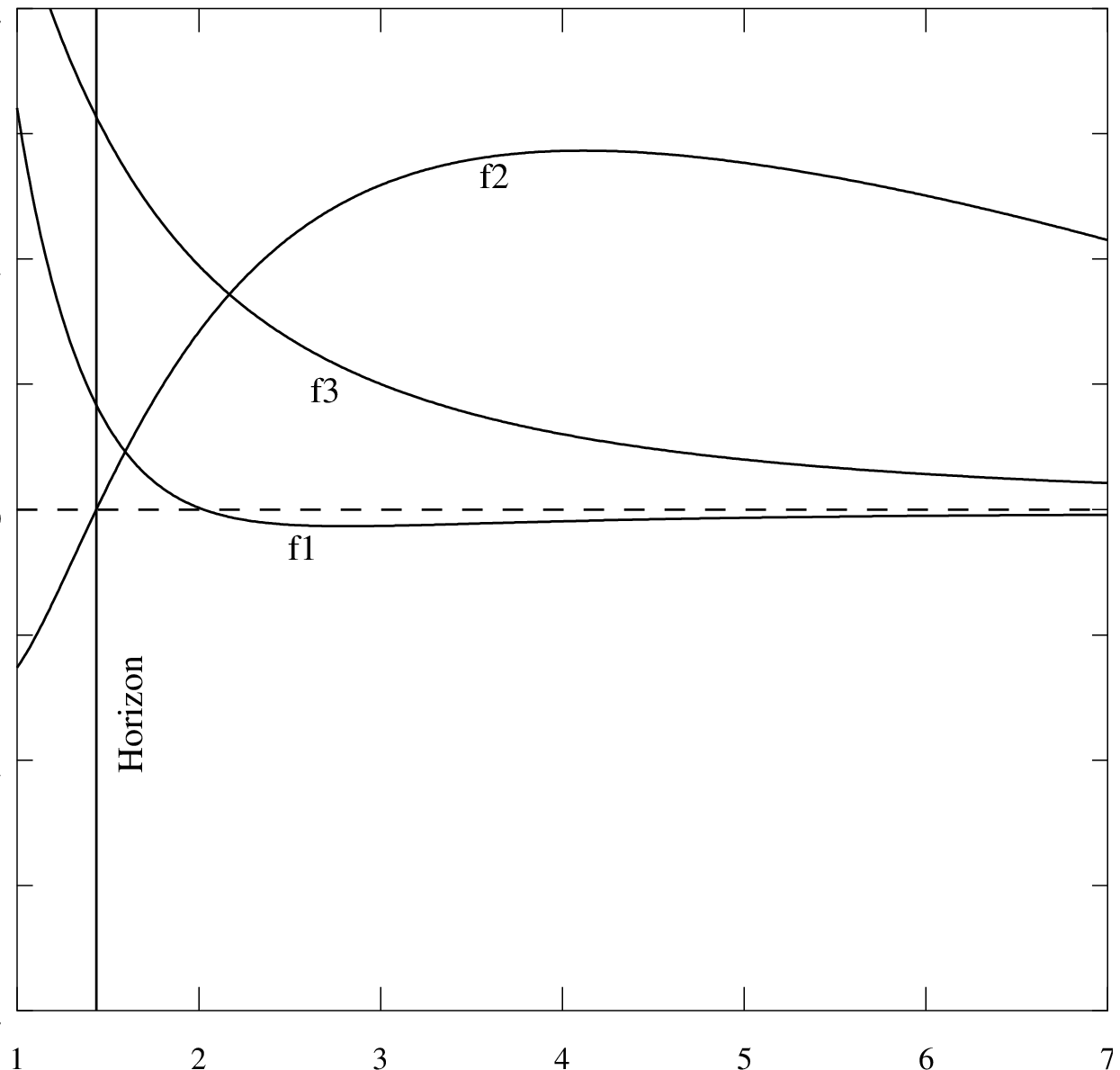}
\end{center}
\singlespace \noindent {\footnotesize Fig. 3:
Same as Fig. 1, but for $a = 0.9$.
}

\middlespace It is well known that the Kerr parameter `$a$' in the
stationary case is interpreted through the asymptotic boundary
condition, as defining the rotation of the source with respect to
distant fixed frame. The concepts of inertial forces, the
centrifugal and Coriolis, as known, do depend on the existence of
a far away inertial frame in a fixed background. The inertial
frame dragging as envisaged in the Kerr geometry is indeed
considered as a true Machian effect (Brill (1994)). We have seen
above how this effect influences the particle in curved geometry.
As the 3-velocity of the test particle does depend upon the local
physics, one can clearly see from the above
example the inherent aspects of Mach's principle in general relativity.\\

\un{Acknowledgement}\\

It is a pleasure to thank Ewald Muller, Toni Font, John Miller and
Marek Abramowicz for discussions at various stages of this work and
Subharti Ray for the help provided in making the plots.
\newpage
{\bf References}\\

\singlespace
 Abramowicz M.A., Carter B. and
Lasota, J.P. (1988), {\it Gen. Rel. \& Grav.} {\bf 20}, 1173.\\

Abramowcicz M.A. and Prasanna A.R. (1990) {\it Mon. Not. R. Astr.
Soc.} {\bf 245}, 720.\\

Abramowicz M.A. (1993) {\it Sci. Am.} {\bf 268}, 74.\\

Abramowicz M.A., Nurowski P. and Wex N. (1993) {\it Class. \& Qu.
Gr.} {\bf 10}, L183.\\

Brill D.R. (1993) {\it Mach's Principle}, Proceedings of the
T\"ubingen Conference (1993), Einstein Studies Vol. 6, Eds. J.
Barbour and H. Pfister, pp. 208-213, Birkha\"user, Boston.\\

Gupta A., Iyer S. and Prasanna A.R. (1996) {\it Class. \& Qu. Gr.}
{\bf 13}, 2675.\\

Iyer S. and Prasanna A.R. (1993) {\it Class. \& Qu. Gr.} {\bf 10},
L13.\\

Hasse W. and Perlick V. (2001) {\it gr-qc} 0108002.\\

Heyl J.S. (2000) {\it Ap. J. (Letters)} {\bf 542}, L45.\\

York J.W. Jr. (1983) `{\it Gravitational Radiation}', ed. N.
Deruelle and T. Piran (North Holland) 175.

\end{document}